\documentclass[aps,prl,twocolumn,groupedaddress]{revtex4-2}

\usepackage{graphicx}
\usepackage{amsmath}

\begin{document}
\title{Observation of a New $X(3872)$ Production Process $e^+e^-\to\omega X(3872)$}

\author{
\begin{small}
\begin{center}
M.~Ablikim$^{1}$, M.~N.~Achasov$^{12,b}$, P.~Adlarson$^{72}$, M.~Albrecht$^{4}$, R.~Aliberti$^{33}$, A.~Amoroso$^{71A,71C}$, M.~R.~An$^{37}$, Q.~An$^{68,55}$, Y.~Bai$^{54}$, O.~Bakina$^{34}$
% ...
\\
\vspace{0.2cm}
(BESIII Collaboration)\\
\vspace{0.2cm} {\it
$^{1}$ Institute of High Energy Physics, Beijing 100049, People's Republic of China\\
% ...
$^{4}$ Bochum  Ruhr-University, D-44780 Bochum, Germany\\
% ...
$^{12}$ G.I. Budker Institute of Nuclear Physics SB RAS (BINP), Novosibirsk 630090, Russia\\
% ...
$^{33}$ Johannes Gutenberg University of Mainz, Johann-Joachim-Becher-Weg 45, D-55099 Mainz, Germany\\
$^{34}$ Joint Institute for Nuclear Research, 141980 Dubna, Moscow region, Russia\\
% ...
$^{37}$ Liaoning Normal University, Dalian 116029, People's Republic of China\\
% ...
$^{54}$ Southeast University, Nanjing 211100, People's Republic of China\\
$^{55}$ State Key Laboratory of Particle Detection and Electronics, Beijing 100049, Hefei 230026, People's Republic of China\\
% ...
$^{68}$ University of Science and Technology of China, Hefei 230026, People's Republic of China\\
% ...
$^{71}$ University of Turin and INFN, (A)University of Turin, I-10125, Turin, Italy; (B)University of Eastern Piedmont, I-15121, Alessandria, Italy; (C)INFN, I-10125, Turin, Italy\\
$^{72}$ Uppsala University, Box 516, SE-75120 Uppsala, Sweden\\
% ...
$^{b}$ Also at the Novosibirsk State University, Novosibirsk, 630090, Russia\\
% ...
}
\end{center}
\vspace{0.4cm}
\end{small}
}

\date{December 14, 2022}

\begin{abstract}
Using $4.7~\rm fb^{-1}$ of $e^+e^-$ collision data at center-of-mass
energies from 4.661 to 4.951 GeV 
collected by the BESIII detector
at the BEPCII collider, we observe the $X(3872)$ production process $e^{+}e^{-}\to\omega X(3872)$
for the first time. 
The significance is $7.8\sigma$, including both the statistical
and systematic uncertainties.
The $e^+e^-\to\omega X(3872)$ Born cross section
and the corresponding upper limit at 90\% confidence level at each energy point are reported.
The line shape of the cross section indicates that the $\omega
X(3872)$ signals may be from the decays of some non-trivial
structures.
\end{abstract}.

\maketitle

A number of experimentally observed quarkonium-like states 
do not fit within the 
conventional quarkonium spectrum and are thus 
popular candidates for exotic hadrons.
As the first experimentally observed quarkonium-like state in this category,
the $X(3872)$ was found 
by Belle in the decay $B^{\pm}\to K^{\pm}\pi^+\pi^-J/\psi$
in 2003~\cite{Belle:2003nnu}.  It was subsequently confirmed by other experiments
~\cite{BaBar:2004oro,CDF:2003cab,D0:2004zmu}.
After two decades of studies, its resonance parameters and 
quantum numbers are well measured. The mass and width 
are determined to be $M=3871.65\pm0.06$ MeV/$c^2$ and $\varGamma=1.19\pm0.21$ MeV, and 
the spin, parity, and $C$-parity quantum numbers are $J^{PC}=1^{++}$~\cite{PDG2022,LHCb:2013kgk,LHCb:2020fvo,LHCb:2020xds}.
The nature of this particle, however, is still not well understood.
Due to the proximity of its mass to the $D^{*0}\bar{D^0}+c.c.$
mass threshold, it is 
conjectured to have a large $D^{*0}\bar{D^0}+c.c.$ molecular component~\cite{intx2}.
The $2^3P_1$ conventional charmonium state $\chi_{c1}(2P)$ with 
$J^{PC}=1^{++}$ is another possible 
interpretation.

In addition to 
exploring the $X(3872)$ via its decays, 
studying its production mechanisms is another way to 
investigate its internal structure. 
Besides its production in $B$ meson decays, 
the $X(3872)$ was also observed in the process $e^+e^-\to\gamma X(3872)$
at BESIII~\cite{BESIII:2013fnz}.
According to the analysis of the line shape of the 
$e^+e^-\to\gamma X(3872)$ cross section,
the $X(3872)$ is produced through the radiative transition 
$Y(4230)\to\gamma X(3872)$.
Large production rates are also observed in prompt production in $pp$
and $p\bar{p}$ collisions~\cite{D0:2004zmu,lhcpp1,lhcpp2,lhcpp3}, 
with rates comparable to the production rates for conventional charmonium states,
which suggests the $X(3872)$ may include a 
conventional charmonium $\chi_{c1}(2P)$ component.
Therefore, searching for new production mechanisms will provide vital information 
to unravel the mysterious nature of the $X(3872)$. 
The hadronic transitions to the spin-triplet charmonium 
states $\chi_{cJ}(1P)$ via $e^+e^-\to\omega\chi_{cJ}(1P)$ $(J=0,1,2)$
have been observed at BESIII~\cite{Beschicj1,Beschicj2,Beschicj3}. 
If the $X(3872)$ contains a component of the excited
spin-triplet charmonium state $\chi_{c1}(2P)$, 
the process $e^+e^-\to\omega X(3872)$ could exist. As the center-of-mass energy
($\sqrt{s}$) threshold to produce $\omega X(3872)$
is about 4.654 GeV, 
the $e^+e^-$ annihilation 
data samples taken at BESIII above this production threshold offer 
an excellent opportunity to search for this process.

In this Letter, we report the first observation of 
the new $X(3872)$ production process $e^{+}e^{-}\to\omega X(3872)$.
The significance is $7.5\sigma$, which includes both the statistical and systematic
uncertainties.
We use
$4.7~\rm fb^{-1}$ of 
$e^+e^-$ annihilation data at $\sqrt{s}$ from 4.661 to 4.951 GeV collected by the BESIII detector. 
We reconstruct the signal process using the decays
$X(3872)\to\pi^+\pi^- J/\psi$, $J/\psi\to \ell^+\ell^-$ ($\ell=e,\mu$),
$\omega\to\pi^+\pi^-\pi^0$, and $\pi^0\to\gamma\gamma$.

The BESIII detector~\cite{BES} has an effective geometrical acceptance of 93\% of $4\pi$.
A helium-based main drift chamber (MDC) in a 1 T magnetic field measures
the momentum and the energy loss (d$E$/d$x$) of charged particles.
The resolution of the momentum is about 0.5\% at 1~GeV/$c$,
and the resolution of the d$E$/d$x$ is better than 6\%.
An electromagnetic calorimeter (EMC) is used 
to measure energies and positions with an energy resolution of
2.5\% in the barrel and 5.0\% in the end caps for 1.0~GeV photons.
A time-of-flight system with a time resolution of 80~ps (110~ps) in
the barrel (end cap) is used for particle identification together with d$E$/d$x$.
A muon chamber (MUC) based on resistive plate chambers with 2~cm position
resolution provides information for muon identification.

Monte Carlo (MC) samples, simulated using {\sc geant4}-based~\cite{geant} software,
are used to optimize the selection criteria, determine
the detection efficiency and study the potential backgrounds. 
The inclusive MC samples, which include open-charm hadronic processes, continuum processes, and the effects due to initial-state-radiation (ISR), are generated with {\sc kkmc}~\cite{KKMC} in conjunction with {\sc evtgen}~\cite{evtgen}. 
The signal MC samples, $e^{+}e^{-}\to\omega X(3872)$, $X(3872)\to\rho^0 J/\psi$, $\rho^0\to\pi^+\pi^-$, 
$J/\psi\to \ell^+\ell^-$ ($\ell=e,\mu$), $\omega\to\pi^+\pi^-\pi^0$, and $\pi^0\to\gamma\gamma$, 
are generated 
with the Final State Radiation (FSR) simulated by {\sc photos}\cite{photos}.
Each track is required to have its point of closest approach
to the beamline within 1~cm in the radial direction and
within 10 cm from the interaction point along the beam direction, and to lie
within the polar-angle coverage of the MDC, $|\cos\theta|<0.93$, in the
laboratory frame.
Photons are reconstructed from isolated showers in the EMC, which are
at least 10 degrees away from any track.
The EMC energy is at least 25~MeV in the barrel ($|\cos\theta|<0.80$) and 50~MeV
in the end-caps ($0.86<|\cos\theta|<0.92$).
In order to suppress
electronic noise and energy deposits unrelated to the event, the EMC
time $t$ of the photon candidate must be in coincidence with collision
events in the range $0\leq t\leq700~\rm{ns}$.

The final state of the signal process includes six charged 
tracks (a lepton pair and four charged pions) and two photons.
In order to improve the detection efficiency, candidate events with five or 
six reconstructed charged tracks are both retained.
Since the leptons from the $J/\psi$ decay have higher momentum than the pions, 
the momentum information is used to separate the leptons from pions.
Two charged particles with momenta greater than 1.0 GeV/$c$ and opposite charges
are identified as the lepton pair from the $J/\psi$ decay, and the remaining tracks 
are each assigned a pion hypothesis. 
Electrons and muons are discriminated by requiring
their deposited energies in the EMC to be 
greater than 0.8~GeV and less than 0.4~GeV, respectively.
In order to suppress the continuum background, at least one muon
from $J/\psi\to \mu^+\mu^-$ needs to penetrate more than three layers 
of the MUC. 
As the pions are either from the $X(3872)$ or $\omega$ and
the combinatorial background
shows smooth distributions in the signal regions according to a study with 
signal MC, all possible combinations are retained.

For the candidate events with six charged particles, the net charge is required to be zero
and at least one photon is required. Instead of 
reconstructing the two photons from a $\pi^0$ decay directly, 
a kinematic fit is applied to 
constrain the recoiling invariant mass of the four pions and the lepton 
pair against the initial $e^+e^-$ collision system to 
the known $\pi^0$ mass~\cite{PDG2022}. The four-momentum of the non-reconstructed $\pi^0$ 
is calculated from 
the kinematic fit. The $\chi^2$ of the one-constraint kinematic 
fit is required to be less than 15. 
The selection criteria are optimized by maximizing $S/\sqrt{S+B}$, where 
the number of signal events ($S$) 
is determined with the signal MC sample
based on the yield observed with an unoptimized selection
in data, and the background ($B$) 
is estimated with an inclusive MC sample.
After applying these requirements, a clear signal peak is seen in
the distribution of the lepton pair invariant mass $M(\ell^+\ell^-)$.
The $J/\psi$ mass window is set to be $3.07 < M(\ell^+\ell^-) < 3.13$~GeV/$c^2$, 
and the sideband regions are defined as $2.96 < M(\ell^+\ell^-) < 3.05$~GeV/$c^2$
and $3.15 < M(\ell^+\ell^-) < 3.24$~GeV/$c^2$ with 
a width of three times the $J/\psi$ mass window to 
estimate the non-$J/\psi$ backgrounds.
The main backgrounds are processes including 
an $\eta$ or $\psi(2S)$ in the final states, {\it e.g.}, 
$e^+e^-\to(\gamma_{\rm ISR})\pi^+\pi^-\psi(2S)$, $\psi(2S)\to\pi^+\pi^- J/\psi$ or 
$\eta J/\psi$.
These backgrounds are reduced 
by requiring all charged track combinations have
$M(\pi^+\pi^-\pi^0)$ and $M(\pi^+\pi^-J/\psi)$  
outside the $\eta$ and $\psi(2S)$ 
mass windows [0.52, 0.58] GeV/$c^2$ and [3.680, 3.692] GeV/$c^2$, 
respectively. The efficiency varies from 18.1\% to 22.0\% at different energy points.
To improve the resolution, 
$M(\pi^+\pi^-J/\psi)$ is defined as $M(\pi^+\pi^-\ell^+\ell^-)-M(\ell^+\ell^-)+m(J/\psi)$,
where $m(J/\psi)$ is the known $J/\psi$ mass~\cite{PDG2022}. 

For signal candidates with one undetected charged pion, 
only five charged tracks are reconstructed.
In this case,
the net charge must be $\pm1$, and at least two photons must be found. 
A two-constraint kinematic fit is applied to these events. 
The invariant mass of the two photons $M(\gamma\gamma)$ is constrained 
to the known $\pi^0$ mass~\cite{PDG2022}, and  
the recoiling mass of the five tracks and the $\pi^0$ 
against the $e^+e^-$ initial collision momentum is constrained 
to the known $\pi^+$ mass~\cite{PDG2022}. If there are more than two photons in an event, 
the combination with the minimum $\chi^2$ from the kinematic fit is chosen. 
The $\chi^2$ is required to be less than 25.
The main backgrounds after placing these requirements are similar to the six-track case, and  
are suppressed using the same criteria. 
The inclusive MC sample, which is five times larger than the data sample, is used to study 
the backgrounds for the five- and six-track cases~\cite{topo}. 
The remaining backgrounds are mainly from continuum processes, and no peaking background is observed. 
The efficiency varies from 5.7\% to 7.1\% at different energy points for the five-track case.

\begin{figure}
\includegraphics[scale=0.21]{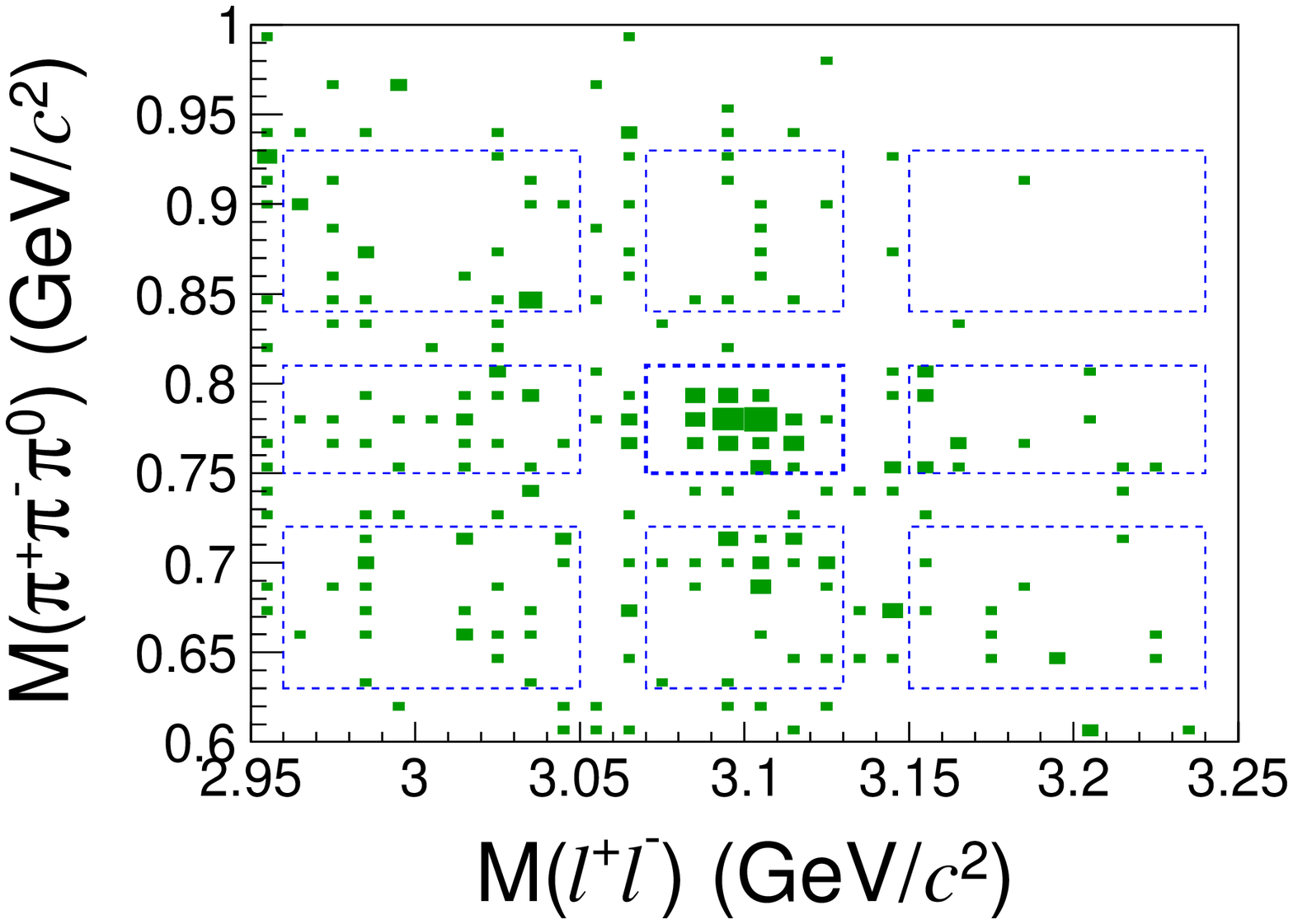}
\includegraphics[scale=0.21]{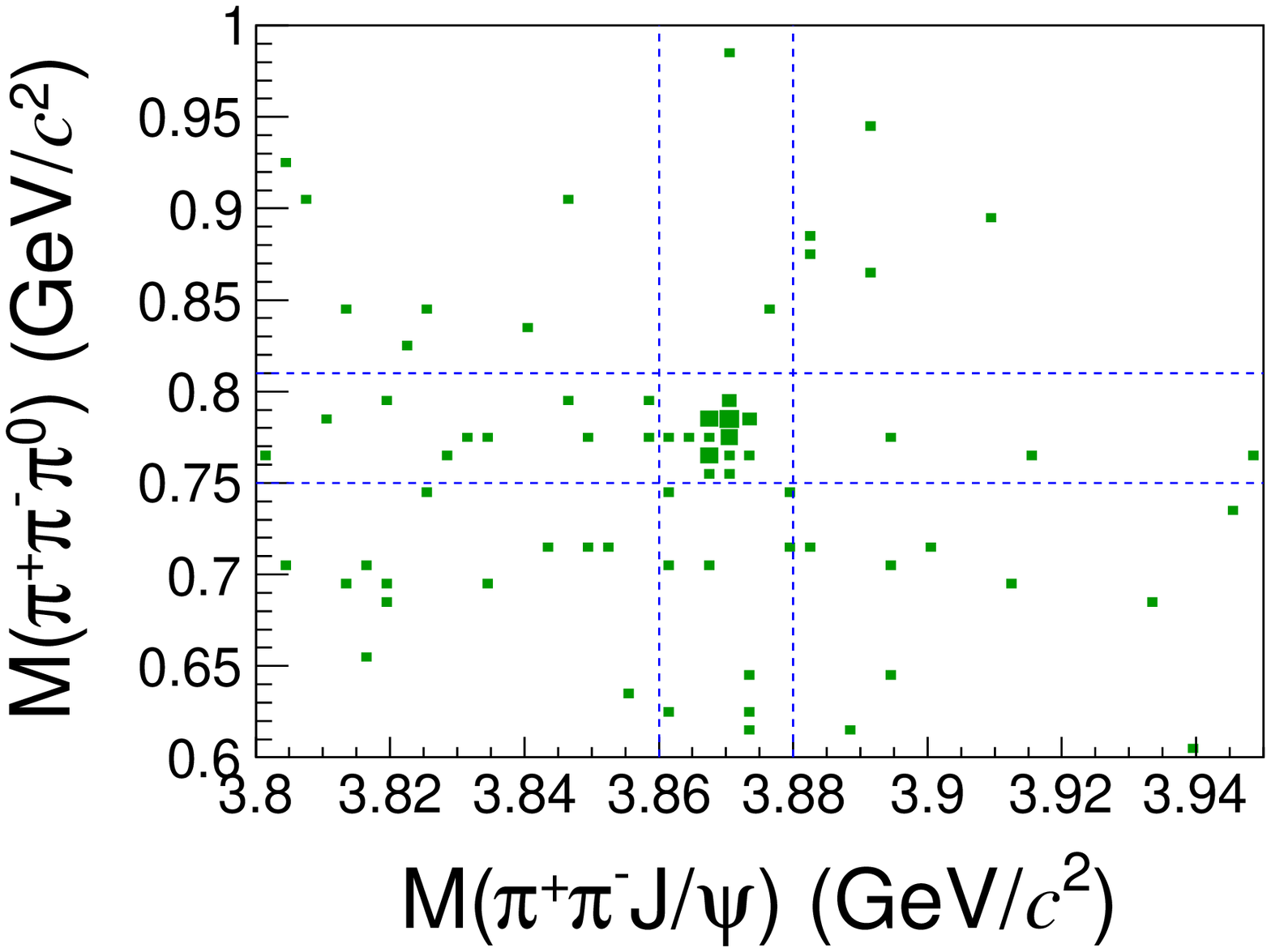}
\caption{Distributions of $M(\pi^+\pi^-\pi^0)$ versus $M(\ell^+\ell^-)$ (left) and $M(\pi^+\pi^-\pi^0)$ versus $M(\pi^+\pi^-J/\psi)$ (right) from the data samples at $\sqrt{s}=4.661-4.951$ GeV.
The central dashed box in the left plot indicates 
the $\omega$ and $J/\psi$ signal regions, and the other boxes show 
the two-dimensional $\omega$ and $J/\psi$ sideband regions. 
The dashed lines in the right plot 
denote the $\omega$ and $X(3872)$
signal regions after imposing the $J/\psi$ selection.}
\label{fig:fit_4660_6950_novan_2d}
\end{figure}

\begin{figure}
\includegraphics[scale=0.35]{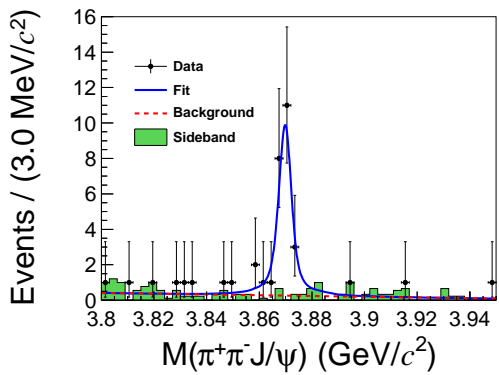}
\caption{Fit to the $M(\pi^+\pi^-J/\psi)$ distribution.
The points with error bars are data, the solid curve is the fit 
result, the dashed line is the background component, and the filled 
histogram represents events from the $\omega$ and $J/\psi$ two-dimensional sidebands.}
\label{fig:fit_4660_6950_novang}
\end{figure}

After applying all the selection requirements, 
Fig.~\ref{fig:fit_4660_6950_novan_2d} shows 
the distribution of $M(\pi^+\pi^-\pi^0)$ versus $M(\ell^+\ell^-)$ and 
$M(\pi^+\pi^-\pi^0)$ versus $M(\pi^+\pi^-J/\psi)$ with the 
combinations of the five- and six-track candidates from the data samples taken at $\sqrt{s}=4.661-4.951$ GeV. 
The presence of the $e^{+}e^{-}\to\omega X(3872)$ signals can
be seen around the intersection of the $\omega$
and $X(3872)$ signal regions.
The $\omega$ candidates are required to have $M(\pi^+\pi^-\pi^0)$ within the $\omega$ mass window
[0.75, 0.81] GeV/$c^2$, and the $\omega$ sideband regions are defined as [0.63, 0.72]~GeV/$c^2$ 
and [0.84, 0.93]~GeV/$c^2$, which are three times the signal region size. 
The sidebands are used to estimate the non-$\omega$ backgrounds. Figure~\ref{fig:fit_4660_6950_novang}
shows the $M(\pi^+\pi^-J/\psi)$ distribution after imposing 
the $\omega$ signal selection. 
A clear peak is seen around the $X(3872)$ signal region.
The two-dimensional $\omega$ and $J/\psi$ 
sidebands are used to check the backgrounds in the $X(3872)$ region
as shown in Fig.~\ref{fig:fit_4660_6950_novang}, which illustrates
a flat distribution.

To determine the $X(3872)$ signal yield and mass, an unbinned maximum
likelihood fit is performed. The signal shape is determined by the signal MC sample with 
an input mass ($m_{\rm input}$) of 3871.7 MeV/$c^2$~\cite{PDG2022}.
The process 
$e^+e^-\to\gamma_{\rm ISR}\pi^+\pi^-\psi(2S)$, $\psi(2S)\to\pi^+\pi^-J/\psi$
is used to calibrate 
the discrepancies in mass ($\Delta m$) and resolution ($\Delta \sigma$) between data and the MC simulation
which are found to be
$-0.5\pm0.2$~MeV/$c^2$ and $0.8\pm0.3$~MeV, respectively. 
The signal MC shape is convolved with a Gaussian function $G(m_{g},\sigma_{g})$ where $m_g$ is free 
and $\sigma_g$ is fixed to $\Delta \sigma$ in the fit.
The background is described with a linear function with free parameters. 
The fit result is shown in Fig.~\ref{fig:fit_4660_6950_novang}. 
The signal yield is $24.6\pm5.3$, and $m_{g}=-2.0\pm0.7$ MeV/$c^2$.
Then, the $X(3872)$ mass is measured to be $m_{x}$ = $m_{\rm input}$ + 
$m_g - \Delta m$ = 3870.2 $\pm$ 0.7 MeV/$c^2$, where the error is statistical only.
Two additional background models are checked in the fit. 
One is to firstly fit the lower mass region [3.8, 3.85] GeV/$c^2$ with 
a linear function and the upper mass region [3.89, 3.95] GeV/$c^2$ with a 
flat distribution, respectively, and then extract the distribution of the background 
in the signal region with a linear interpolation between the two fitting functions, and another is a 3rd-order polynomial function.
A point-to-point dissimilarity method is applied to test the 
goodness-of-fit~\cite{gof}. The yielded $p$-values corresponding to the three background models 
are 0.93, 0.94, and 0.82, respectively, which indicate good agreement between the data and the fit results.

The significance is estimated by comparing the difference of 
log-likelihood values $\Delta(-2 \ln \mathcal{L})$ with and without the $X(3872)$ signal 
in the fit, and taking the change in the number of degrees of freedom $\Delta ndf$ into account.
Various fit schemes e.g. different fitting ranges and background models
are applied to extract the significance.
We take the obtained smallest value $7.8\sigma$ as 
the significance in consideration of the systematic uncertainties.
The systematic uncertainty of the mass measurement is mainly caused by 
the $\Delta m$ error (0.2 MeV/$c^2$).
The uncertainty due to the mass shift in simulation is assigned to be 0.1 MeV/$c^2$ 
according to study the signal MC sample. 
The uncertainties from the fit are 
estimated by changing different fit schemes, and obtained to be 0.1 MeV/$c^2$ in total. 
The total uncertainty of the $X(3872)$ mass measurement is 0.3 MeV/$c^2$ by summing
all these uncertainties assuming they are independent.

The $e^+e^-\to\omega X(3872)$ Born cross section 
is calculated by
\begin{equation}
\sigma^{\rm B}=\frac{N_{\rm sig}}{\mathcal{L}_{\rm{int}}\mathcal{B}_{1}(\epsilon_{ee}\mathcal{B}_{ ee}+\epsilon_{\mu\mu}\mathcal{B}_{\mu\mu})(1+\delta)\frac{1}{|1-\Pi|^{2}}},
\label{cal1}
\end{equation}
where
$N_{\rm sig}$ is the number of 
signal events; 
$\mathcal{L}_{\rm{int}}$ is the 
integrated luminosity; 
$\epsilon_{ee}$ and $\epsilon_{\mu\mu}$ 
are the detection efficiencies 
of the electron and muon modes, respectively; 
$\mathcal{B}_{ee}$ and $\mathcal{B}_{\mu\mu}$ are 
the branching fractions of $J/\psi\to e^+e^-$ and 
$J/\psi\to\mu^+\mu^-$, respectively;  
$\mathcal{B}_{\rm 1}$ is the product of the branching fractions of 
$\omega\to\pi^+\pi^-\pi^0$ and $X(3872)\to \pi^+\pi^-J/\psi$;
$(1+\delta)$ is the ISR correction factor 
obtained by using a QED calculation~\cite{QED-Delt} iteratively by taking 
the measured cross section in this analysis
as an input; and 
$\frac{1}{|1-\Pi|^{2}}=1.055$ is the vacuum polarization 
factor taken from QED with an accuracy of 0.05\%~\cite{Vacuum-Delt}.
All branching fractions are  taken from Ref. \cite{PDG2022}.

Due to the limited statistics, 
the signal yield ($N_{\rm sig}$) at each energy point is determined
by counting the number of events in the $X(3872)$
signal region [3.86, 3.88]~GeV/$c^2$. 
The background has been subtracted, which is estimated by using 
the $X(3872)$ sidebands [3.81, 3.84] GeV/$c^2$ and [3.91, 3.94] GeV/$c^2$.
Only the [3.81, 3.84] GeV/$c^2$ sideband region is used at $\sqrt{s}=4.661$~GeV 
since $M(\pi^+\pi^-J/\psi)$ has a maximum allowed value of 3.914~GeV/$c^2$ at that energy.
The measured $N_{\rm sig}$ and $\rm \sigma^B$
at each energy point are listed in Table~\ref{sum_meas}.
The statistical significance and the
upper limits of $\rm \sigma^B$ ($\sigma^{\rm B}_{\rm up}$) at the 90\% confidence level
at various energy points
are calculated using a frequentist method 
with an unbounded profile likelihood treatment 
by assuming the numbers of observed events in the $X(3872)$ 
signal and sideband regions follow a Poisson distribution~\cite{rolke}.

\begin{table*}
\caption{Summary of the 
integrated luminosities ($\mathcal L_{\rm int}$), the signal yields ($N_{\rm sig}$), the product of average efficiency ($\epsilon=(\epsilon_{ee}+\epsilon_{\mu\mu})/2$) and the ISR correction factor ($1+\delta$), the obtained $e^+e^-\to \omega X(3872)$ Born cross section or its upper limit ($\sigma^{\rm B}_{\rm up}$), and statistical significances at various energy points. The first uncertainties of the Born cross sections are statistical, the second are systematic, 
and the third are the uncertainties caused by the branching fraction of $X(3872)\to \pi^+\pi^-J/\psi$.
}
\begin{tabular}{ccccccc}
\hline\hline
$\sqrt{s}~(\rm GeV)$&$\mathcal{L}_{\rm{int}} (\rm pb^{-1})$ &$ N_{\rm sig}$&$\epsilon(1+\delta)$ (\%)&$\rm \sigma^{B} (pb)$&$\rm \sigma^{B}_{up} (pb)$& Significance \\
\hline
4.661&  529.63&$0.33^{+1.36}_{-0.33}$&28.3&$0.5^{+2.1}_{-0.5}\pm0.1\pm0.2$&5.6&-\\
4.682& 1669.31&$8.00^{+3.34}_{-2.68}$&24.6&$4.6^{+1.9}_{-1.5}\pm0.4\pm1.5$&11.5&3.4$\sigma$ \\
4.699& 536.45 &$0.00^{+0.95}_{-0.00}$&27.0&$0.0^{+1.6}_{-0.0}\pm0.0\pm0.0$&3.3&-\\
4.740& 164.27 &$1.67^{+1.77}_{-1.10}$&21.8&$10.9^{+11.6}_{-7.2}\pm1.0\pm3.5$&40.6&1.0$\sigma$\\ 
4.750& 367.21 &$5.00^{+2.58}_{-1.92}$&22.4&$14.2^{+7.4}_{-5.5}\pm1.4\pm4.5$&38.2&3.1$\sigma$\\
4.781& 512.78 &$1.00^{+1.36}_{-0.70}$&31.6&$1.5^{+2.0}_{-1.0}\pm0.2\pm0.5$&6.5&0.7$\sigma$\\ 
4.843& 527.29 &$4.67^{+2.58}_{-1.92}$&26.7&$7.8^{+4.3}_{-3.2}\pm0.7\pm2.5$&21.1&2.6$\sigma$\\
4.918& 208.11 &$1.00^{+1.36}_{-0.70}$&22.6&$5.0^{+6.8}_{-3.5}\pm0.4\pm1.6$&21.7&0.7$\sigma$\\ 
4.951& 160.37 &$0.00^{+0.95}_{-0.00}$&20.4&$0.0^{+6.8}_{-0.0}\pm0.0\pm0.0$&14.7&-\\
\hline\hline
\label{sum_meas}
\end{tabular}
\end{table*}

\begin{table}[h]
\small
\caption{\label{tab:table4}
Relative systematic uncertainties (in \%) 
in the Born cross section measurements at various energy points. 
The uncertainties caused 
by the integrated luminosity ($\sigma_{\mathcal{L}}$), 
the detection efficiency ($\sigma_{\epsilon}$), the ISR correction factor ($\sigma_{\rm ISR}$), the method of 
signal extraction ($\sigma_{\rm sig}$), and their sum in quadrature ($\sigma_{\rm sum}$) are listed. }
\begin{tabular}{lccccc}
\hline\hline
$\sqrt{s}$ (GeV) & $\sigma_{\mathcal{L}}$ &$\sigma_{\epsilon}$ &$\sigma_{\rm ISR}$ & $\sigma_{\rm sig}$ &$\sigma_{\rm sum}$\\
\hline
4.661&1.0&8.1&5.0&1.6&9.7\\
4.682&1.0&8.1&2.3&1.6&8.6\\
4.699&1.0&8.1&12.0&1.6&14.6\\
4.740&1.0&8.1&4.3&1.6&9.4\\
4.750&1.0&8.2&5.4&1.6&10.0\\
4.781&1.0&8.3&12.2&1.6&14.9\\
4.843&1.0&8.3&1.4&1.6&8.6\\
4.918&1.0&8.4&1.2&1.6&8.7\\
4.951&1.0&8.5&0.5&1.6&8.7\\
\hline\hline
\label{sys_tem}
\end{tabular}
\end{table}

The systematic uncertainties of the Born cross section measurement mainly 
originate from the detection efficiency,
the ISR correction factor, 
the method of signal extraction, the integrated luminosity, and the branching 
fraction of $X(3872)\to \pi^+\pi^-J/\psi$.
The sources of the uncertainty from the detection efficiency include the tracking, 
the photon reconstruction, the kinematic fit, the $J/\psi$ mass window, 
the muon selection, and the signal generation model. 
The systematic uncertainty due to tracking 
is estimated with the process $e^+e^-\to\pi^+\pi^- J/\psi$ to be 1.0\% per track~\cite{track}.
The uncertainty of the photon reconstruction efficiency is assigned to be 1.0\% per photon from the study of the process $J/\psi\to\rho^0\pi^0$~\cite{photon}.
The uncertainties 
caused by the kinematic fit, 
the $J/\psi$ mass selection, and 
the muon selection with the MUC are studied 
with the control sample of $e^+e^-\to\gamma_{\rm ISR}\pi^+\pi^- \psi(2S)$,
$\psi(2S)\to\pi^+\pi^- J/\psi$. 
The corresponding uncertainties are 2.8\%, 4.2\%, and 1.3\%, respectively. 
The efficiencies are calculated with the signal
MC samples generated with a phase space model which is flat 
in the distributions of $\omega$ and $\rho^0$ helicity angles. 
The uncertainty caused by the generation model and the $\eta/\psi(2S)$ veto
is estimated 
by varying the distributions of the $\omega$ and $\rho^0$ helicity 
angles $\theta$ to be $1\pm \rm cos^2\theta$. The uncertainties are found to be (0.1-1.7)\%.
The signal yield at each energy point is obtained with the counting method; a 1.6\% uncertainty is 
assigned comparing to that obtained with an alternative fit method at $\sqrt{s}=4.684$ GeV.
The uncertainty due to the ISR correction factor
is estimated by scaling the initial input observed line-shape within 
one statistical uncertainty, and the relative difference of the efficiency 
compared to the nominal scheme is taken as the uncertainty, 
which varies from 1.4\% to 12.0\% at different energy points. The integrated luminosity is measured with the 
Bhabha scattering process with an uncertainty 
of 1.0\%~\cite{lum}. The total systematic uncertainty at each energy point is obtained 
by adding all these systematic uncertainties in quadrature. The systematic uncertainties discussed above are summarized in Table~\ref{sys_tem}. The uncertainty caused by the branching fraction of $X(3872)\to \pi^+\pi^-J/\psi$
is 31.6\%~\cite{PDG2022}, which is listed as a separate uncertainty in the Born cross section.

In summary, based on data samples  
at $\sqrt{s}=4.661-4.951$ GeV with 
a total integrated luminosity of $4.7~\rm fb^{-1}$ collected by the BESIII detector, a new $X(3872)$ production process
$e^+e^-\to\omega X(3872)$ is observed for
the first time. The 
significance is $7.8\sigma$, including the statistical and systematic uncertainties.
The measured $X(3872)$ mass is $3870.2\pm0.7\pm0.3$ MeV/$c^2$.
The $e^+e^-\to\omega X(3872)$ Born cross section
and the corresponding upper limit at the 90\% confidence level
at each energy point are reported. 
The line shape of the cross section indicates that the 
observed $\omega X(3872)$ signals may be from decays of some non-trivial structures.
The production mechanisms of the $X(3872)$ provide 
crucial information about its properties.
The observation of a new production
process $e^+e^-\to\omega X(3872)$,
combined with the observation of the $X(3872)$
in other production mechanisms,
offers an additional window into the
composition of the $X(3872)$.

\begin{acknowledgments}
The BESIII collaboration thanks the staff of BEPCII and the IHEP computing center for their strong support. This work is supported in part by National Key R\&D Program of China under Contracts Nos. 2020YFA0406300, 2020YFA0406400; National Natural Science Foundation of China (NSFC) under Contracts Nos. 11805090, 12147214, 11635010, 11735014, 11835012, 11935015, 11935016, 11935018, 11961141012, 12022510, 12025502, 12035009, 12035013, 12192260, 12192261, 12192262, 12192263, 12192264, 12192265; Outstanding Research Cultivation Program of Liaoning Normal University No. 21GDL004; the Chinese Academy of Sciences (CAS) Large-Scale Scientific Facility Program; Joint Large-Scale Scientific Facility Funds of the NSFC and CAS under Contract No. U1832207; the CAS Center for Excellence in Particle Physics 
(CCEPP); 100 Talents Program of CAS; The Institute of Nuclear and Particle Physics (INPAC) and Shanghai Key Laboratory for Particle Physics and Cosmology; ERC under Contract No. 758462; European Union's Horizon 2020 research and innovation programme under Marie Sklodowska-Curie grant agreement under Contract No. 894790; German Research Foundation DFG under Contracts Nos. 443159800, Collaborative Research Center CRC 1044, GRK 2149; Istituto Nazionale di Fisica Nucleare, Italy; Ministry of Development of Turkey under Contract No. DPT2006K-120470; National Science and Technology fund; National Science Research and Innovation Fund (NSRF) via the Program Management Unit for Human Resources \& Institutional Development, Research and Innovation under Contract No. B16F640076; Olle Engkvist Foundation under Contract No. 200-0605; STFC (United Kingdom); Suranaree University of Technology (SUT), Thailand Science Research and Innovation (TSRI), and National Science Research and Innovation Fund (NSRF) under Contract No. 160355; The Royal Society, UK under Contracts Nos. DH140054, DH160214; The Swedish Research Council; U. S. Department of Energy under Contract No. DE-FG02-05ER41374.
\end{acknowledgments}

\bibliographystyle{plain}

\end{document}